\documentclass[12pt,tightenlines,
    onecolumn,
    preprintnumbers,
    amsmath,
    amssymb,
    prd,
    showpacs,
    showkeys
]{revtex4}
\usepackage{color}
\usepackage{graphicx}%
\usepackage{amsmath}
\usepackage[normalem,normalbf]{ulem}
\usepackage{amssymb}
\usepackage{bm}
\usepackage{enumerate}

\newcommand{\nn}{\nonumber}

\begin{document}
\title{Casimir Energy of a Spherical Shell in $\kappa-$Minkowski Spacetime}
\author{Hyeong-Chan Kim$^1$\thanks{Kavli Institute for Theoretical Physics China(KITPC) at the Chinese Academy of Sciences, Beijing, China.}}
\email{hckim@phya.yonsei.ac.kr}
\author{Chaiho Rim$^2$}
\email{rim@chonbuk.ac.kr}
\author{Jae Hyung Yee$^1$}%
\email{jhyee@yonsei.ac.kr}
\affiliation{$^1$ Department of Physics, Yonsei University,
Seoul 120-749, Republic of Korea\\
$^2$ Department of Physics and Research Institute
of Physics and Chemistry, Chonbuk National University,
Jeonju 561-756, Korea.
}%
\bigskip
\begin{abstract}
We study the Casimir energy of a spherical shell 
of radius $a$ in $\kappa$-Minkowski spacetime 
for a complex field with an asymmetric ordering 
and obtain the energy up to $O(1/\kappa^2)$.
We show that the vacuum breaks particle 
and anti-particle symmetry
if one requires the spectra to be consistent 
with the blackbody radiation
at the commutative limit.
\end{abstract}
\maketitle

\section{Introduction}
Since Casimir first predicted
that the quantum fluctuation
of the electromagnetic field would produce an
attractive force between two infinite parallel plates
in vacuum~\cite{casimir},
the Casimir energy has been found to depend on
the geometry of the system:
The Casimir force is repulsive
for a spherical  geometry~\cite{boyer2}.
The Casimir effect has attracted much
attention experimentally and theoretically~\cite{milton}.
The effect has now been measured within
about the one percent error range 
and at distances
down to tens of nanometers
for parallel plates as reported in Ref.~4.
The idea has been applied to a wide range of phenomena,
from explaining the amazing ability
of a geko to walk across the ceilings~\cite{autumn},
to a possible way of understanding the Hawking
radiation~\cite{boyers}, and to
stabilizing the radion field
for resolving the hierarchy problem
in the brane-world scenarios~\cite{garriga}.

On the other hand, 
at short distances of the Planck length scale,
the spacetime itself may change its form
due to the quantum gravity effect.
Especially, $\kappa$-deformed Poincar\'{e}  algebra
(KPA) is introduced~\cite{kappaP}.
Here, the four momenta commute with each other,
but the boost relation is deformed,
where $\kappa$ has the role of the deformation parameter.
In this dual picture, when $\kappa$ approaches infinity,
the deformed Poincar\'{e} symmetry reduces to the commutative limit, the ordinary Poincar\'{e} symmetry.
The deformed realization implies a
deformed special relativity
that results in a change of the group velocity of the photon.
In this respect, doubly special relativity~\cite{doubly}
is closely related with this KPA~\cite{doubly-kappa}
and the deformation parameter $\kappa$
reflect the Planck-scale physics.

After the appearance of the KPA,
it was soon realized that
the dual picture of the KPA results 
in a non-commuting spacetime~\cite{majid}.
This non-commuting spacetime is called
the $\kappa$-Minkowski spacetime (KMST),
in which  the rotational symmetry is preserved,
but time and space coordinates
do not commute each other:
\begin{equation}
~[\hat x^0, \hat x^i] = \frac{i}{\kappa} \hat x^i\,,\qquad
 ~[\hat x^i, \hat  x^j]=0\,,\qquad i,j=1,2,3\,.
\end{equation}
The differential structure of the KMST of 4 spacetime
dimensions is not realized in 4-dimensional spacetime
but needs to be constructed
in 5-dimensional spacetime~\cite{sitarz,gonera}.
If the corresponding derivative is realized
in 5-dimensional momentum space
${\cal P}_A$ ($A=0,1,2,3,5)$,
then the derivatives satisfy
the  4-dimensional De Sitter space
\begin{equation}
({\cal P}_0)^2-\sum_{i=1}^3({\cal P}_i)^2 -({\cal P}_5)^2 = -\kappa^2\,.
\end{equation}
It is noted that $P_5$ is invariant under
the KPA; therefore, 
if one requires the physical system to preserve
the $\kappa$-deformed Poincar\'{e}  symmetry (KPS),
then one can restrict oneself to 4-dimensional
spacetime, including the derivatives.

Based on this differential structure,
the scalar field theory has been
formulated~\cite{kosinski,KRY-field}.
The $\kappa$-deformation was extended
to a curved space with a $\kappa$-Robertson-Walker metric
and was applied to the cosmic microwave
background radiation~\cite{kim}.
The effect of the $\kappa$-deformation on the blackbody
radiation has been studied recently in 
Ref.~17.

Still, KMST is not understood well,
and  an interacting (field) theory,
including gauge symmetry, needs more elaboration 
because many particle properties show a  non-local nature
(See Refs.~18 
and 19 
and references therein).
For a systematic study, one needs to look into KMST
and see if KMST field theory allows a reliable vacuum
in which a particle picture
can be constructed from the vacuum.
In this sense, the Casimir energy
can provide a useful check on the nature
of the vacuum.

As noted above, KMST field theory
is constructed on the dual space of KMST
through KPS.
To do this, one defines a field variable in momentum space as
\begin{equation}
\phi(x) \equiv
\int \frac{d^4 p}{(2\pi)^4}  \,
e^{-i p \cdot  x}\,\varphi (p)\,.
\end{equation}
Here, all the coordinate variables
and momenta are treated as commuting variables.
Instead, the non-commuting nature of KMST is encoded
in the $\ast$-product between field variables.
The product of an exponential element is required to
satisfy the composition rule \cite{FGN}
\begin{equation}
 e^{-i p \cdot  x}\ast e^{-i q
\cdot x} = e^{-i v (p, q) \cdot x }\,,
\end{equation}
where we will adopt in this paper
the composition law
corresponding to the asymmetric ordering
\begin{equation}
v (p, q)
= ( p^0+q^0, {\bf p} e^{-q^0/\kappa}+ {\bf q}) \,.
\end{equation}
In this approach, the spacetime variables $x^\mu$ are
treated as commuting with each other
and the effect of the original spacetime non-commutativity is given in terms of the homomorphism of the field variables
through the $\ast$-product.
One can, thus, avoid various conceptual difficulties
of spacetime geometry,
which arises from the non-commutating nature
of the spacetime.

The KPS in the dual picture is the guiding principle to construct the field theory and is applied to the free scalar action explicitly
in Ref.~15. 
The free analogue of massive complex scalar theory is
given as
\begin{equation}
\label{action-coor}
S= \int d^4 x \, \phi^c(x)\ast
 \left[-\partial_\mu \ast   \, \partial^\mu \ast
  - m^2 \right] \, \phi(x)\,.
\end{equation}
$\phi^{c}(x)$ is the conjugate of the scalar field
\begin{equation}
\phi^{c}(x)
\equiv \int \frac{d^4 p}{(2\pi)^4}
e^{3 p^0/\kappa} \, e^{i \tilde
p \cdot  x}\,\varphi^\dagger(p)
=\int \frac{d^4 p}{(2\pi)^4}
e^{-ip \cdot x}\,\,\varphi^\dagger(- \tilde p )\,,
\end{equation}
where $ \tilde p^0 = p^0$
and $\tilde {\bf p} =  e^{p^0/\kappa}\,{\bf p}$,
and $\varphi^\dagger(p)$
denotes the ordinary complex conjugate of
$\varphi(p)$ in momentum space.
The measure factor $e^{3p^0/\kappa}$ and $\tilde p^\mu$
are needed to satisfy the KPS.

In momentum space, the action in Eq.~(\ref {action-coor})
is given as
\begin{equation}
S= \int \frac{ d^4 p}{(2\pi)^4} \, e^{3 p^0/\kappa}
\varphi^\dagger(p) \Delta_F ^{-1}(p)
 \, \varphi(p) \,.
\end{equation}
The integration measure is KPS invariant, and
the ``Feynman propagator'' is given as
\begin{equation}
\label{feynman}
\Delta_F ^{-1}(p) = M^2_\kappa(p)
\left(1+\frac{M^2_\kappa(p)}{4\kappa^2} \right)
 -m^2  + i\epsilon\,,
\end{equation}
where $M_\kappa^2(p)$ is the first Casimir invariant
\begin{equation}
\label{eq:casimir-asm-inv}
 M_\kappa^2(p)=\left(2 \kappa \sinh \frac{p_0}{2
\kappa }\right)^2- {\bf p}^2 e^{p_0/\kappa}
\end{equation}
and a small positive real number, $\epsilon$,
is added to avoid the singularity on the
real axis of $p^0$.
Explicitly, the Feynman propagator
is given as
\begin{equation}
\label{propagator}
\Delta_F ^{-1}(p) =   \frac{\kappa^2}4 e^{2p_0/\kappa}
    (e^{-p_0/\kappa} - \alpha_+) (e^{-p_0/\kappa} +\alpha_+)
   (e^{-p_0/\kappa} - \alpha_-) (e^{-p_0/\kappa} + \alpha_-)\,,
\end{equation}
where
\[
\alpha_\pm  =
 \sqrt{1+ \frac{m^2-i\epsilon}{\kappa^2}}
\pm \sqrt{ \frac{m^2+\textbf{p}^2 -i\epsilon}{\kappa^2}}\,.
\]
The Feynman propagator has the periodic property
\begin{equation}
 \Delta_F ^{-1}(p_0+ i \kappa \pi, {\bf p} )
=\Delta_F ^{-1}(p_0, {\bf p})\,,
\end{equation}
and, thus, possesses an infinite number of
poles on the complex plane of $p^0$.
Nevertheless, the real poles
provide a stable particle
and an anti-particle dispersion relation, and
one can study the physical effects
of the modified dispersion relation
by simply ignoring the unstable modes
because the unstable modes 
decay very quickly after the Planck time has passed.
In this spirit, the blackbody spectra
has been investigated in Ref.~17 
for the massless scalar theory.
The massless dispersion relation is given as
\begin{equation}
\label{masslessmode}
\omega_{\bf p}^{(+)}= -\kappa \ln (1-| {\bf
p}|/\kappa) \,,\quad
\omega_{\bf p}^{(-)} = \kappa \ln (1+| {\bf
p}|/\kappa)\,,
\end{equation}
where $\omega_{\bf p}^+$ corresponds to the particle
dispersion relation and
$\omega_{\bf p}^-$ to the anti-particle's.
It is demonstrated that the thermal fluctuation
of the particle is different from that of the anti-particle.
The Stephan-Boltzmann law is modified at the order of $O(1/\kappa)$.
However, due to the different thermal behaviors 
of the particle and the antiparticle,
the $O(1/\kappa)$ effects cancel each other, 
and the Stephan-Boltzmann law 
is left with an $O(1/\kappa^2)$ correction
when both the particle and the anti-particle are present.

The effect of the $\kappa$-deformation 
for the case of two infinite parallel plates 
on the Casimir effect has 
been studied in Ref.~21 
using the real pole in Eq.~(\ref{feynman}).
The deformed effect was found to be the order 1/$\kappa^2$.
A similar effect was also shown 
at the order of 1/$\kappa^2$  
in Ref.~22 
and 23 
when  a different dispersion relation was used.
The different dispersion relation 
corresponds to
a different realization of KPS 
even though KMST is the same.
The two investigations demonstrate
that a different realization of KSP
may result in a different correction
to the physical effect.

In this paper, we calculate the Casimir energy
for a sphere of radius $a$ and
study the particle and the antiparticle contributions
to the vacuum energy.
In the commutative spacetime,
the Casimir energy is positive
for a spherical boundary.
It would be interesting
how the $\kappa$-deformation alters this Casimir energy
and how it affects the vacuum.
In Sec.~\ref{sec:reg},
we illustrate the computational
procedure for the Casimir energy
with a spherical boundary in the KMST,
closely following
that of Refs.~\cite{boyer2,nesterenko,hagen}.

In Sec.~\ref{sec:Ec2} and Sec.~\ref{sec:Ec3},
we compute the Casimir energy
of the anti-particle mode, $\omega_{\bf p}^{(-)} $,
in Eq.~(\ref{masslessmode}).
In Sec.~\ref{sec:Ec2}
the Casimir energy is calculated
without the measure factor in the momentum space,
where the momentum variable is
treated as a mere mode-counting parameter.
In Sec.~\ref{sec:Ec3}, the Casimir energy is computed,
including the measure factor.
In Sec.~\ref{sec:discussion} we summarize
the Casimir energies given
in Sec.~\ref{sec:Ec2} and Sec.~\ref{sec:Ec3}
and compare the results with the energy
of the particle mode,
$\omega_{\bf p}^{(+)}$, in Eq.~(\ref{masslessmode}).
We discuss the particle and the anti-particle symmetries
of the vacuum and 
present the ordering effect on the symmetry of the
vacuum.
Some detailed calculations are given in the appendices:
The calculation of the divergent part, ${\cal E}_0$, 
is given in the App.~\ref{sec:appA}, of the
$O(1/\kappa)$ correction in App.~\ref{sec:appB},
and of the $O(1/\kappa^2)$ correction 
in App.~\ref{sec:appC}.

\section{Casimir energy of a spherical shell} \label{sec:reg}
In this section, we present an idea
on how to calculate the Casimir energy
of a massless scalar field
in $\kappa$-Minkowski spacetime
for a spherical shell of radius $a$.
The Casimir energy is the zero point vacuum energy of
massless scalar fields.
The massless modes are 
are given in Eq.~(\ref{masslessmode}).
We note that the particle mode
$\omega^{+}_{\bf p}$  is defined when
$ |{\bf p}| < \kappa$ whereas
the anti-particle mode
$\omega^{-}_{\bf p}$  is defined for all momentum.
The Feynman propagator in Eq.~(\ref{propagator})
turns out to provide an additional pole 
on the real axis
when $ |{\bf p}| > \kappa$,
\begin{equation}
\label{eq:hm}
p_0^{(3)}=  -\kappa \ln (| {\bf p}|/\kappa -1)\,,
\end{equation}
in addition to the two modes
\begin{equation}
p_0^{(+)}= \omega^{(+)}_{\bf p} \,, \qquad
p_0^{(-)}=- \omega^{(-)}_{\bf p}\,.
\end{equation}

Thus, all three real modes contribute
to the Casimir energy
\begin{equation}
E_c
=  \frac {\hbar} 2 \sum_{i=+,-,3}\sum_{{\bf p}}
\omega^{(i)}_{\bf p}
\end{equation}
where $\omega^{(3)}_{\bf p}$ refers to the
mode related with $p_0^{(3)}$
whose relation is not fixed yet.
The difficulty lies in that the value of
$p_0^{(3)}$ ranges from + infinity to - infinity
in contrast with $p_0^{(+)}$ and $p_0^{(-)}$.
Thus, we will divide the Casimir energy into two parts,
$E_c = E_c{(A)} + E_c{(P)}$, where
\begin{equation}
E_c^{(A)} =  \frac {\hbar} 2 \sum_{{\bf p}}
\omega^{(-)}_{\bf p} \,,\qquad
E_c^{(P)}=  \frac {\hbar} 2 \sum_{i=+,3}\sum_{{\bf p}}
\omega^{(i)}_{\bf p}
\end{equation}
so that the momentum ranges from $-\infty$ to $\infty$.
In the momentum configuration,
the vacuum energy will be represented
as
\begin{eqnarray}
\label{casimirenergy}
E_c^{(A)}
&=&  \frac {\hbar} 2 \int \frac{d^3 \bf p}{(2\pi)^3}
\,\, \omega^{(-)}_{\bf p} \,
e^{\alpha p_0^{(-)}/{\kappa} }\,,
\\
E_c^{(P)}
&=&  \frac {\hbar} 2 \Big(  \int_{|{\bf p}|< \kappa}  \frac{d^3 \bf p}{(2\pi)^3}
  \,\, \omega^{(+)}_{\bf p} \,
e^{\alpha p_0^{(+)}/{\kappa} }
  +
 \int_{|{\bf p}|> \kappa}  \frac{d^3 \bf p}{(2\pi)^3}
   \,\,\omega^{(3)}_{\bf p} \,
e^{\alpha p_0^{(3)}/{\kappa} }
   \Big)\,,
\end{eqnarray}
where $\alpha=3$
from the $\kappa$-Poincar\'{e} invariant measure.

Since there is an ambiguity in $ E_c^{(P)} $,
we will consider  $E_c^{(A)}$ first.
One can find the momentum mode contribution in a
spherical shell for
$p_0^{(-)}= - \hbar \omega^{(-)}_{\bf p} $
by using the wave equation in coordinate space:
\begin{eqnarray} \label{eq:wave:asym}
- \nabla^2 \psi(\vec x, t)= \lambda^2  \psi(\vec x, t)\,,
\end{eqnarray}
where ${\bf p}^2 = \hbar^2 \lambda^2$.
It is noted that
in this dual (momentum space) picture,
the KMST effect is entirely encoded
in the dispersion relation,
Eq.~(\ref{masslessmode}),
through the $\ast$-product,
the spacetime coordinates are 
treated as commuting variables,
and, thus, the ordinary quantum mechanical tool
can be employed without conceptual difficulty,
which chiefly arises
from the spacetime non-commutativity.
Especially, one can separate the time and the space
coordinates in the wave function
$\psi({\bf x}, t) =
\phi({\bf x} ) \, e^{-i \omega_{\bf p} t}$
and to arrive at the eigenvalue equation,
by using spherical symmetry,
\begin{eqnarray} \label{eq:eigen}
\left\{r^2 \frac{d^2}{dr^2}+ 2r \frac{d}{dr}+\left[\lambda^2 r^2
    -l(l+1)\right]\right\}\phi_l(r) = 0\,.
\end{eqnarray}
There is a $(2l+1)$-fold degeneracy
in the  eigenvalues $\lambda$.
Explicitly, the solution is given by the spherical Bessel functions:
\begin{equation} \label{sol}
\phi(r) = \Big\{ \begin{array}{ll}
 j_l(\lambda r) & \quad \mbox{ for } r<a \\
A_l j_l(\lambda r)+ B_l n_l(\lambda r) &\quad \mbox{ for } r>a \,,
\end{array}
\end{equation}
where the regularity is imposed at $r=0$ and
$A_l$ and $B_l$ are constants to be determined by prescribing the correct
asymptotic behavior at large $r$.

At this stage, one can follow the usual trick to 
impose the boundary conditions~\cite{boyer2,nesterenko,hagen}.
At $r=a$, one imposes the Dirichlet boundary condition
\begin{equation}\label{eq:bc}
j_l(\lambda a) = 0\,, \qquad
A_l j_l(\lambda a) + B_l n_l(\lambda a) = 0\,.
\end{equation}
In addition, to find the asymptotic behavior at $r \to \infty$,
one may conveniently regularize the exterior modes
by enclosing the entire system within another concentric sphere of radius
$R\gg a$.
The boundary condition at large $R$,
$ A_l j_l(\lambda R)+B_l n_l (\lambda R)=0$,
gives the phase
\begin{eqnarray} \label{delta:l}
\tan \delta_l\equiv \frac{B_l}{A_l} = \tan (\lambda R-\frac{l\pi}{2}) .
\end{eqnarray}

To accommodate the boundary condition at $r=a$
for the modes inside and outside,
one may define an analytic function
\begin{eqnarray} \label{eq:fl}
\tilde f_l(z) =  f_l^{(1)}(z) f_l^{(2)}(z)\,,
\end{eqnarray}
with $f_l^{(1)}(z) = j_l(z)$ and
$f_l^{(2)}(z) =  j_l(z)+ \tan\delta_l(z) n_l(z) \,,$
where $z=\lambda r$.
Then, the boundary condition in Eq.~(\ref{eq:bc}) 
is written as $\tilde f_l (z_n) =0$,
where $z_n=\lambda_n a$ and $ \lambda_n$ 
is the quantized value of $\lambda$
due to the spherical boundary.
One uses the Cauchy theorem 
to write the sum of analytic functions
\begin{eqnarray} \label{Residue}
\sum_i \zeta(x_i) = \frac{1}{2\pi i}\oint _C \zeta(z)
\frac{\tilde f'_l(z)}{\tilde f_l(z)} dz
=  \frac{1}{2\pi i}\oint _C dz \zeta (z) \frac{d}{dz}\log \tilde f_l(z),
\end{eqnarray}
where $x_i$'s are isolated zeros of  $\tilde f_l(z)$ within a closed contour $C$.
The Casimir energy given by the sum of the vacuum modes,
\begin{eqnarray} \label{eq:Ec0}
E_c^{(A)}
= \sum_{l=0}^\infty\sum_{n=1}^\infty \left(l+\frac{1}{2}\right)
    \omega (a; z_n)
    e^{-{\alpha \omega  (a;z_n)}/{\kappa } }\,,
\end{eqnarray}
is then written as
\begin{eqnarray}
\label{eq:preEc}
E_c^{pre} (a)  =\frac{1 }{2\pi i} \sum_{l=0}^\infty
    \left(l+\frac{1}{2}\right)\oint_C d z \,
    e^{-\frac{\alpha \omega(a;z)}{\kappa } }\, \omega(a;z)
    \frac{d}{dz} \ln  f_l(z) \,,
\end{eqnarray}
where  $ \omega(a;z) = \kappa \ln (1+z/(\kappa a))$
and
\begin{equation}
f_l (z) =2 \, z^2 \tilde f_l (z)
=  2 \, z^2  f_l^{(1)}(z) f_l^{(2)}(z) \,.
\end{equation}
Here, we introduced a pole at $z=0$
without changing the value of the integral in 
Eq.~(\ref{eq:preEc}), noting that $\omega(a;0)=0$.
This freedom allows one to replace  
$\tilde f_l (z)$ by $f_l (z) $.

The expression of the Casimir energy 
in Eq.~(\ref{eq:preEc})
needs a few comments.
In  general, 
the expression does not converge when
summing  over the modes. It is noted that the
high-frequency mode grows rapidly
as the momentum becomes large.
Thus, in general, one has to regularize the expression first
and find a mean to find the finite contribution.
To regularize, we conveniently
introduce an infinitesimal positive
parameter $\sigma$ and write the energy as
\begin{eqnarray} \label{eq:Ec}
E_{c}^{reg} (a)
= \frac{1 }{2\pi i} \sum_{l=0}^\infty
    \left(l+\frac{1}{2}\right)\oint_C d z \,
    e^{-\sigma z- \frac{\alpha \omega(a;z)}{\kappa } }\,
    \omega(a;z) \frac{d}{dz} \ln f_l(z) \,.
\end{eqnarray}
The factor $e^{-\sigma z}$ plays the role of 
a cutoff
and suppresses the high-frequency contributions
to the Casimir energy.
In our case, however, the presence of the measure factor
provides a natural cut-off effect already.
Nonetheless, we will carry the $\sigma$
for future convenience when the case of $\alpha=0$
is considered for comparison.

Next, the amount of vacuum energy 
without a spherical boundary 
is subtracted from this expression  
to obtain the net
vacuum energy due to a sphere of radius $a$.
To do this, one calculates the energy
for a large sphere of radius $ \eta R $
($\eta$ is a finite number on the order of 1
so that $a< \eta R < R$)
and subtracts the result from the expression 
in Eq.~(\ref{eq:Ec}):
\begin{equation}
E^{(A)}_c (a)
= \lim_{ R \to \infty, \sigma \to 0 }
\Big( E_{c}^{reg}  (a) - E_{c}^{reg}  (\eta R)  \Big)\,.
\end{equation}

\begin{figure}[htbp]
\begin{center}
\includegraphics[width=.45\linewidth,origin=tl]{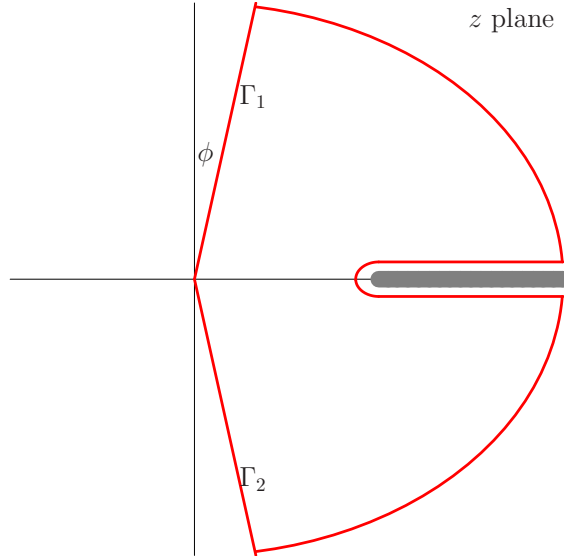}
\end{center}
\caption{Contour for the integration.}
\label{fig:contour}
\end{figure}
To compute the Casimir energy in Eq.~(\ref{eq:Ec}),
we take the  contour $C$ for the
integration, as shown in Fig.~\ref{fig:contour},
which can be conveniently broken into three parts:
a circular segment $C_\Lambda$, and
two straight line segments $\Gamma_1$ and
$\Gamma_2$.
Since the $\Gamma$ contours are oriented
at a nonzero angle $\phi$
with respect to the imaginary axis, it follows that the
contribution to $C_\Lambda$ (especially when $\alpha=0$)
is bounded by $\exp(-\sigma \Lambda \sin\phi)$,
where $\Lambda$ is the radius of the circular arc.
Since the logarithm in the Casimir energy grows
at most algebraically,
it follows that the contribution to $C_\Lambda$ vanishes
exponentially in the
limit of large $\Lambda$, provided that $\phi\neq 0$.

Along $\Gamma_1$, setting the coordinate $z=i y e^{-i \phi}$ with $y$ real leads to 
\begin{eqnarray} \label{delta l}
\tan \delta_l
= \tan ( i y e^{- i\phi} \frac{R}{a} -\frac{l\pi}{2}) \to i  
\end{eqnarray}
for sufficiently large $R \gg a$
and
\begin{eqnarray}
f_l^{(1)}(z) &=&
\sqrt{\frac{\pi}{2z} } J_\nu (z)
= e^{i \pi \nu/2} \sqrt{\frac{\pi}{2z}}
        I_\nu (y e^{-i \phi}),\\
f_l^{(2)}(z) &=&
\sqrt{\frac{\pi}{2z} } H^{(1)}_\nu (z)
= -e^{-i \pi \nu/2} \Big(\frac{2i} \pi \Big) \sqrt{\frac{\pi}{2 z}}K_\nu (y e^{-i\phi})\,,
\end{eqnarray}
with $\nu=l+1/2$.
Thus, on $\Gamma_1$, 
one has $f_l(z) = \lambda_\nu (ye^{-i\phi}) $,
where
\begin{equation}
\lambda_\nu (y) =2 y  I_{\nu}(y) K_{\nu}(y)\,.
\end{equation}
The contribution from $\Gamma_2$
is the complex conjugate of the $\Gamma_1$ contribution.
 This gives the Casimir energy
\begin{eqnarray}\label{eq:Ec2}
E_c^{(A)} (a) &=&  
\lim_{R \to \infty, \sigma \to 0, \phi \to 0}
\sum_l \Big(E_{l}^{reg}(a)
-  E_{l}^{reg}(\eta R) \Big)\,,
\end{eqnarray}
where
\begin{equation}
\label{eq:regE}
E_{l}^{reg} (r) = \frac{\kappa\nu}{\pi}
 \Re \int_0^\infty dy\, e^{- i \sigma ye^{-i \phi}}
i g(r, i y e^{-i\phi}) \frac{d}{dy} \log \lambda_\nu (ye^{-i\phi})\,,
\end{equation}
with
\begin{equation}
\label{g-function}
g(r,z)  =
\frac{\log\left(1+\frac{z}{\kappa r}\right)}
{\left(1+\frac{z}{\kappa r}\right)^\alpha } \,.
\end{equation}

To sum up the angular momentum modes,
one may conveniently use the large-$\nu$ behavior 
of the Bessel function.
After shifting $y \to \nu y$ in the integration,
\begin{equation} \label{eq:dI}
E_l^{reg}(r)
=  \frac{\kappa\nu}{\pi}  \Re\int_0^\infty dy\,
 i e^{- i \sigma \nu ye^{-i \phi}} g(r, i \nu y e^{-i \phi})
 \frac{d}{dy} \log {\lambda_\nu( \nu ye^{-i\phi})}\,,
\end{equation}
one uses the large-order series expansion of 
the Bessel function~\cite{abramwitz}
for $\nu \gg 1 $:
\begin{eqnarray} \label{eq:largeorder}
\log\lambda_\nu(\nu y)=
    \sum_{n=0}^\infty\frac{{q}_n(y)}{\nu^{2n}}\,.
\end{eqnarray}
$q_n(y)$ is a function of $O(y^{-2n})$ for large $y$,
whose
explicit forms  for $n=0,1,2$ are given in 
Eq.~(\ref{eq:q12}).
This manipulation results in the Casimir energy
\begin{equation}
\label{eq:Casimir-largeorder}
E_c(a) = \sum_{n=0}^\infty \Big( {\cal E}_n (\sigma, a)
- {\cal E}_n (\sigma, \eta R) \Big)\,,
\end{equation}
where
\begin{eqnarray}
{\cal E}_n (\sigma, r)  &=&   \sum_l  \frac{\kappa}{\pi\nu^{2n-1}}
\Re\int_0^\infty dy\,
i  \Big(e^{- i \sigma \nu ye^{-i \phi}}\, g(r, i \nu y e^{-i \phi})
)\Big)\frac{d}{dy}
        q_n(y e^{-i \phi})
\end{eqnarray}
with the limits $R \to \infty, \sigma \to 0$, and 
$\phi \to 0$  being taken at the end.

This decomposition of the Casimir energy
in Eq.~(\ref{eq:Casimir-largeorder}) is useful
in taking care of the divergent structure
in the $1/\kappa$ expansion.
First, one can be convinced that ${\cal E}_0 (a)$
vanishes 
because the integration gives only a pure imaginary
contribution,
as shown in Appendix~\ref{sec:appA}.
(A similar conclusion 
can be made using the zeta function
regularization as in Ref.~24.) 
The rest of the terms with  $n \ge 1$ are finite
even when the limits
$\sigma \to 0$ and $\phi \to 0$ are taken
before the summation over $l$ and integration over $y$.
Thus, the finite Casimir energy is simplified  as
\begin{equation}
\label{eq:finalenergyform}
E_c(a) = \sum_{n=1}^\infty \lim_{R \to \infty}
\Big( {\cal E}_n (a) - {\cal E}_n (\eta R) \Big)\,,
\end{equation}
where ${\cal E}_n (r)$ is summed up with angular momentum
contributions
\begin{equation}
\label{eq:Enr}
{\cal E}_n (r) = \frac1r \sum_l  \frac{B_n (\nu, r) }{\nu^{2n-2}}\,;
\qquad B_n (\nu, r)  = \frac1{\pi }
\int_0^\infty dy\, q_n(y )\,\,  G\left(\frac{\nu y}{\kappa r}\right)\,.
\end{equation}
Here, integration by parts is used, and
$G(x)$ is an even function of $x$:
\begin{equation}
\label{eq:Gx}
G (x)=
\left\{
\begin{array}{ll}
\frac{1}{1+x^2} & \mbox {for } \alpha=0 
\vspace{.1cm}\\
\frac{(1- 6x^2+x^4)(1-3 \log \sqrt{1+x^2})- 12 (1-x^2) x\tan^{-1}x}
	{(1+x^2)^4}&\mbox {for } \alpha=3 \,.
    \end{array}
    \right.
\end{equation}

As $\kappa \to \infty$,
$G(\frac {\nu y}{\kappa a} ) \to 1$.
In this commutative limit, one may have $B_n (\nu, r) \to B_n ^{(0)}(\nu, r)$,
\begin{equation}
{\cal E}_n(r) \to {\cal E}_n^{(0)} (r)
= \frac1r \sum_l  \frac{ B_n ^{(0)}(\nu, r)}{\nu^{2n-2}}\,,
\end{equation}
and $E_c (a) \to  E_c^{(0)} (a)$,
whose expression is exactly the same as the one 
given in Ref~27, 
\begin{equation}
\label{eq:comm-casimir}
E_c^{(0)}(a) = \frac{0.002819 }{a}\,.
\end{equation}
Then, the higher-order terms in 1/$\kappa$
are given as
\begin{equation}
\label{eq:higher-En}
\Delta {\cal E}_n (r)
= {\cal E}_n (r) - {\cal E}_n^{(0)} (r)
= \frac1{r} \, 
\sum_l \frac{\Delta B_n(\nu, r)}{ \nu^{2n-2}}\,,
\end{equation}
where
\begin{equation}
\label{eq:Bn-qn}
\Delta B_n(\nu, r)
\equiv B_n(\nu, r) - B_n ^{(0)}(\nu, r)
= \frac{1}{\pi}\int_0^\infty dy
    q_n(y) \,\left[ G \left(\frac{\nu y}{\kappa r}\right)-1\right] \,.
\end{equation}
The correction terms are considered
in the next two sections.

It is obvious from Eqs.~(\ref{eq:Enr}) and
(\ref{eq:Gx}) that  $E_c(a)$ is independent of the sign of $\kappa$. Thus, one may expect
the particle and the anti-particle to give the
same contributions to the Casimir energy. However, there
arises a subtle point due to the presence of the
branch cut in the particle dispersion relation
$\omega^{+}_{\bf p}$. This will be carefully
investigated in the last section.

\section{Casimir energy for the $\alpha=0$ case }\label{sec:Ec2}

Let us consider the $\alpha=0$ case in this section.
This case neglects the KPS invariant measure
in the integration, but
is simpler than the non-zero $\alpha$ case
and provides an  informative structure
in the systematic calculation in the $1/\kappa$ series
expansion.

The higher-order contribution of $ B_n (\nu)$
is given in Eq.~(\ref{eq:Bn-qn}),
whose explicit expression is given as
\begin{equation}
\Delta B_n(\nu, r)
= -\frac{\nu^2 }{\pi (\kappa r)^2}
    \int_0^\infty dy
    \frac{q_n(y) \, y^2 }{1+\frac{\nu^2 y^2}{(\kappa r)^2}} \,.
\end{equation}
The details of the calculations of
$\Delta {\cal E}_1 (r)$ and $\Delta {\cal E}_2 (r)$
are given in Appendix~\ref{sec:appB}.
A large sphere of radius $\eta R$ 
only gives a non-trivial contribution 
to $\Delta {\cal E}_1 (\eta R)$;
$\Delta {\cal E}_{n \ge 2} (\eta R)$  vanishes
as $R \to \infty$.
The finite correction terms
$\Delta {\cal E}_1 (r)$ and $\Delta {\cal E}_2 (r)$
are of the order of $O(1/|\kappa|)$ and  are given as
\begin{eqnarray} \label{eq:sumBC}
&&\Delta {\cal E}_1 (a) - \Delta {\cal E}_1 (\eta R)
=  \frac 1 a \sum_{l=0}^\infty \Big( \Delta B_1(\nu, a)
-\Delta B_1(\nu, \eta R ) \Big)
= -\frac 1 a \left ( \frac{1}{384}\frac{1}{|\kappa| a}
   + O\Big(\frac1{(\kappa a)^{3}}  \Big)\right)  \,,\nn
\\
&&\Delta {\cal E}_2 (a)
= \frac 1 a \sum_{l=0}^\infty \frac{\Delta B_2(\nu,a )}{\nu^2} 
=  \frac 1 a \left (
    \frac{1}{256} \, \frac{1}{|\kappa| a}
    + O\Big(\frac1{(\kappa a)^{3}}  \Big)\right)\,.
\end{eqnarray}
Here, we put the absolute value notation to $\kappa$, 
even though $\kappa$ is positive, 
to emphasize that
$  {\cal E}_n (a) $ is  independent of the sign of
$\kappa$.

The dominant contribution of
$\Delta {\cal E}_{n\ge 3} (a)$
is considered in Appendix \ref{sec:appC}:
\begin{equation}
\Delta  {\cal E}_n (a)
= \frac 1{\pi a} \sum_l \frac {1 } {\nu^{2n-2}}
\int_0^\infty dy \, {q_n(y) } \Big(
{G\left(\frac{\nu y}{\kappa a}\right) - G(0)} \Big)\,,
\end{equation}
and its summation is expressed as
\begin{eqnarray}
\sum_{n \ge 3} \Delta  {\cal E}_n (a)
&=& E_c ^{(2)} (a) + E_c ^{(3)} (a)\,,
\\
E_c ^{(2)} (a)
&=&  -\frac 1{\pi a} \sum_{n \ge 3}\sum_{l} \frac {1 } {\nu^{2n-2}}
\int_0^\infty dy \, {q_n(y) }
 \left(\frac{\nu y}{\kappa a}\right)^2\,,
\nn\\
E_c ^{(3)} (a)
&=&  \frac 1{\pi a} \sum_{n \ge 3}\sum_l \frac {1 } {\nu^{2n-2}}
\int_0^\infty dy \, {q_n(y) }
\frac { (\frac{\nu y}{\kappa a} )^4}{1+ (\frac{\nu y}{\kappa a} )^2 } \,.
\nn
\end{eqnarray}
Noting $q_n (y)= O (y^{-2n})$,
one can confirm that
$E_c ^{(2)} (a)$ and $ E_c ^{(3)} (a)$
are of the orders of $O(1/\kappa^2)$ and $O(1/\kappa^3)$, respectively.
$E_c ^{(2)} (a)$ is calculated with the help of numerics.
The first 10 angular momentum modes are obtained
numerically and are shown to converge to the asymptotic expression for large $l$. This allows one to find the
numerical value accurately, 
whose value is given in Eq.~(\ref{Ec-k2}):
\begin{equation}
E_c^{(2)} = \frac1 a \left(- \frac{0.000545} {(\kappa a)^{2}}\right)\,.
\end{equation}
Combining all the terms, we have
\begin{eqnarray}
E_{c}^{(A)} (a)
= \frac{1}{a}\left(0.002819+\frac{1}{768 (|\kappa| a)}
   -\frac{0.000545}{(\kappa a)^{2}}
   +O\left(\frac 1{\kappa a}\right)^3\right)\,.
\end{eqnarray}
%
\section{Casimir energy for the $\alpha=3$ case} \label{sec:Ec3}

We now take into account the measure effect.
To find $\Delta {\cal E}_n (r) $, 
one needs to take care
of the non-rational function $G(x)$ given in
Eq.~(\ref{eq:Gx}),
\[
G(x)= \frac{(1- 6x^2+x^4)
	  (1-3 \log \sqrt{1+x^2})- 12 (1-x^2) x\tan^{-1}x}{
    (1+x^2)^4}\,.
\]
In this case, $G(x)$ does not allow
easy summation over $l$.
To estimate $\Delta {\cal E}_n (r) $, we note that
$\log (1+x^2)$ and $x\tan^{-1}x$
satisfy $0 \leq \log(1+x^2)\leq x\tan^{-1} x \leq x^2$,
with the equalities holding only at $x=0$.
Thus, one can estimate the range of $G(x)$:
\begin{eqnarray} \label{G:limits}
&&G_{\rm min}(x) \le G(x) \le G_{\rm max}(x) ,\\
&&G_{\rm min}(x) =\frac{2 - 39 x^2 + 2 x^4 - 3 x^6}{2 (1 + x^2)^4},
\nn\\
&&G_{\rm max}(x) =\frac{1 - 6 x^2 + 22 x^4}{(1 + x^2)^4} \,.\nn
\end{eqnarray}

\smallskip

\begin{figure}[htbp]
\begin{center}
 \includegraphics[width=.45\linewidth,origin=tl]{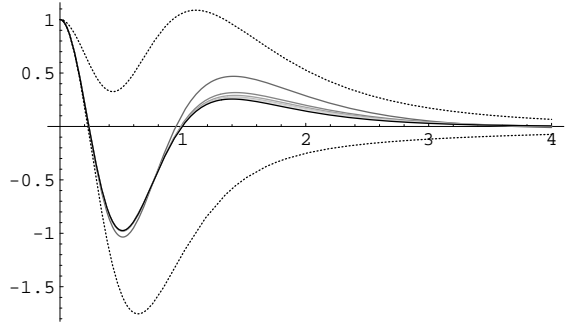}
\end{center}
\caption{$G$ function (black curve). 
The dashed curves denotes $G_{\rm max}$ and $G_{\rm min}$,
respectively. The gray curves denote a systematic 
approximation of $G(x)$ by using Eqs.~(\ref{approx-log}) and 
(\ref{approx-xarctan}).
} \label{fig:G}
\end{figure}

If we get the contribution to the order of $O(1/\kappa)$,
we only have to consider $\Delta {\cal E}_1 (r)$ and
$ \Delta {\cal E}_2 (r)$
becasue $\Delta {\cal E}_{n\ge 3} (r) $
contributes to the
$O(1/\kappa^2)$, as seen in the previous section.
Suppose we use $G_{\rm max}$ to evaluate $O(1/\kappa)$:
\begin{eqnarray}
\nn
  {\cal E}_1 (r)  + {\cal E}_2 (r)
&&=\frac1 r \sum_{l=0}^\infty\frac{1}{\pi}
\int_0^\infty dy \, G_{\rm max}(b y)
\left[ q_1(y)+\nu^{-2} q_2(y)\right] \\
\nn
&&=-\frac 1r \left( \frac{63 \kappa r}{1024}
    -\frac{35 \pi^2}{65536}-\frac{43}{192288\kappa r}
    + O\left(\frac1{\kappa r }\right)^3 \right)\,.
\end{eqnarray}
Subtracting $\mathcal{E}_1(\eta R)$ from $\mathcal{E}_1(a)$,
we find that  the $O(\kappa)$ contribution goes away and
that a 
finite contribution is obtained as $R\rightarrow \infty$.
The $\kappa$-independent term is already contained 
in Eq.~(\ref{eq:comm-casimir}),
and the $1/\kappa$ contribution is obtained as
\begin{equation}
\label{Ec-max}
E_{\rm c\, B1}^{(1)}\equiv
\sum_{i=1}^2 \Big(
\Delta {\cal E}_i (a) - \Delta {\cal E}_i (\eta R) \Big)
=\frac1a \frac{43}{192288\kappa a}
\cong \frac1a \frac{0.00022}{\kappa a} \,,
\end{equation}
which will give a lower bound to $E_{\rm c}^{(1)}(a)$.

Suppose we use $G_{\rm min}$ to evaluate
$ {\cal E}_1 (r)$ and $ {\cal E}_2 (r) $.
We then have
\begin{eqnarray}
\nn
{\cal E}_1 (r)  + {\cal E}_2 (r)
&&= \frac1r
\sum_{l=0}^\infty\frac{1}{\pi}
\int_0^\infty dy \, G_{\rm min}(b y)
\left[ q_1(y)+\nu^{-2} q_2(y)\right]\\
\nn
&&= \frac1r \left(\frac{63 \kappa r}{1024}
    +\frac{35 \pi^2}{65536 }+\frac{133}{12288\kappa r}
    + O\left(\frac1{\kappa r }\right)^3 \right) \,.
\end{eqnarray}
This gives an upper bound to the $1/\kappa$ contribution:
\begin{equation}
\label{Ec-min}
E_{\rm c\,B2}^{(1)}\equiv
\sum_{i=1}^2 \Big(
\Delta {\cal E}_i (a) - \Delta {\cal E}_i (\eta R) \Big)
=\frac1a \left(\frac{133}{12288\kappa a} \right)
\cong \frac1a \left(\frac{0.01082}{\kappa a} \right) \,.
\end{equation}
Comparing the results in Eqs.~(\ref{Ec-max}) 
and (\ref{Ec-min}),
we have lower and upper bounds
on the $1/\kappa$ contribution 
to $E_c^{(1)}(a) $, respectively,
\begin{equation}
E_{\rm c\,B1}^{(1)} < E_{\rm c}^{(1)}(a)
<  E_{\rm c\,B2}^{(1)} \,.
\end{equation}

One may find a good approximate value of
$ E_{\rm c}^{(1)}(a)$
if one finds a good approximation of $G(x)$ in
a quotient form.
To do this, one may approximate $\log(1+x^2)$ as
\begin{eqnarray}
\label{approx-log}
f_1(x)&=&x^2 \,, \\ \nn
f_2(x)& =&\frac{1 + x^2/2}{1 + x^2}\,  f_1(x) \,, \\ \nn
f_3(x)& =& \frac{1 + 2x^2 + 5 x^4/6}{(1 + x^2)^2}\, f_2(x) \,,\\ \nn
f_4(x)& =& \frac{1 + 3x^2 + 3x^4 + 5x^6 /6}{(1 + x^2)^3}\,f_3(x) \,,\\
f_5(x) &=& \frac{1 + 4x^2 + 6x^4 + 4x^6 + 13x^8 /15}{(1 + x^2)^4 }\, f_4(x) \,,\nn
\end{eqnarray}
and $x\tan^{-1}x$ as
\begin{eqnarray} \label{approx-xarctan}
h_1(x)&= &x^2 \,,\\ \nn
h_2(x)& =&  \frac{1 + 2 x^2 /3}{1 + x^2} \,h_1(x)\,,\\ \nn
h_3(x)& =& \frac{1 + 2x^2 + 13 x^4 /15}{(1 + x^2)^2}\,h_2(x)\,, \\ \nn
h_3(x)& =& \frac{1 + 3x^2 + 3x^4 + 277x^6 /315}{(1 + x^2)^3}\,h_3(x) \,,\\
h_5(x) &=&  \frac{1 + 4x^2 + 6x^4 + 4x^6 +  859x^8/945}{(1 + x^2)^4 }\,h_4(x)\,, \nn
\end{eqnarray}
and so on.
The approximate functions $f_n(x)$ and $h_n(x)$ agree with  $\log(1+x^2)$ and $x \tan^{-1} x$, respectively,
up to $O(x^{2n})$ for small $x$.
In addition, one can show that the bounded values of
$\frac{f_n(x)-\log(1+x^2)}{(1+x^2)^4}$ and
$\frac{h_n(x)-x\tan^{-1}x }{(1+x^2)^4}$
improve as $n$ increases for the whole integration range of $x$.
For example,
$\frac{f_n(x)-\log(1+x^2)}{(1+x^2)^4}$ is bounded by 0.0036024 when $n=2$, and
as one uses higher $n$, the  bounded value decreases by around $(1/2)^n$.
The same thing holds for $h_n(x)$.

Using the approximate functions  $f_n(x)$ and $h_n(x)$,
one can integrate and sum over $l$ to get
$$
E_{{\rm c}\, n}^{(1)} = \frac1a \left(\frac{D_n}{\kappa r} \right)\,,
$$
where $D_1=0.009114$, $D_2=0.009094$, $D_3=0.009100$, $D_4=0.009106$, and $D_5=0.009109$.
From these approximated results, we have
\begin{equation}
E_{\rm c}^{(1)}= \frac1a \left(\frac{D}{\kappa r} \right) \,,
\end{equation}
with $D \cong  0.00911$, 
which value is close to the upper bound  $E_{\rm c\, B2}^{(1)}$ in Eq.~(\ref{Ec-min}).

We may find the $O(1/\kappa^2)$ contribution by
summing over
$\mathcal{E}_{n \ge 3}(r)$. 
This contribution
is easily read from Eq.~(\ref{Ec-k2}) 
by using  the coefficient $G_1= -47/2$:
\begin{equation}
\label{Ec-2}
E_c^{(2)} = \frac{0.001713}{\pi a} \,\frac{G_1} {(\kappa a)^{2}}
= \frac1a \left( - \frac{0.01281} {(\kappa a)^{2}}\right) \,.
\end{equation}
From this consideration, we conclude that
the Casimir energy is given by
\begin{eqnarray}
E_{c,\alpha=3}^{(A)}(a)
&=&\frac{1}{a}\left(0.002819+\frac{0.00911}{|\kappa| a}
    -\frac{0.01281}{(\kappa a)^{2}} +O(\frac{1}{(\kappa a)^{3}})\right)\,.
\end{eqnarray}
It is noted that the sign of the first-order term allows the Casimir force to be more repulsive than that of
the commutative result. The first-order term is stronger 
than it is in the case where the measure factor is neglected.

\section{Summary and Discussion}
\label{sec:discussion}

We have evaluated the Casimir energy
in $\kappa$-Minkowski spacetime
when the massless scalar anti-particle
mode satisfies the Dirichlet boundary condition
at a spherical boundary of radius $a$.
The boundary condition is incorporated
using the Cauchy integration.
The scalar theory is used
in the $\ast$-product formalism and
is required to satisfy
the $\kappa$-deformed Poincar\'e symmetry
in momentum space
to avoid the conceptual difficulty
due to the non-commutative nature of the
time and the space coordinates.

The Casimir energy is regulated by the
introduction of a cut-off function
(for the case when the integration measure is neglected),
and a geometry independent term
is subtracted to find the spherical geometric effect.
The Casimir energy  which
respects the $\kappa$-deformed Poincar\'{e} invariance
is given as
\[
E_{c,\,\alpha=3}^{(A)} (a)
=\frac{1}{a}\left(0.002819+\frac{0.00911}{|\kappa|a}
    -\frac{0.01281}{(\kappa a)^{2}}
    +O(\frac{1}{(|\kappa| a)^{3}})\right)\,.
\]
On the other hand,
if one regards the momentum in the integration
of Eq.~(\ref{casimirenergy})
as a mere mode-counting parameter
and neglects the integration measure
(i.e., $\alpha=0$ case),
the Casimir energy is given by
\[
E_{c,\, \alpha=0}^{(A)} (a) = \frac{1}{a}\left(0.002819+\frac{1}{768 (|\kappa| a)}
   -\frac{0.000545}{(\kappa a)^{2}}
   +O\left(\frac 1{|\kappa| a}\right)^3\right)\,.
\]
This shows that the $\kappa$-deformed
Poincar\'{e} invariant measure affects an
physical values such as the Casimir energy.
In addition, the $\kappa$-deformed spacetime
seems to give an 
additional positive contribution at 
long distances and to provide an attractive contribution
at short distances around $\kappa a \cong O(1)$.

The Casimir energy is an even function of $\kappa$
and is independent of the sign of $\kappa$.
On the other hand, $\kappa \to -\kappa$
changes the energy, 
$ \omega_{\bf p}^{(-)}  \to \omega_{\bf p}^{(+)}$,
in Eq.~(\ref{masslessmode}). This seems to suggest 
there is a particle and an
anti-particle symmetry in the vacuum.
However, if one tries to compute the Casimir energy 
by using the positive mode 
$\omega_{\bf p}^{(+)}$,  one encounters
a branch-cut at $\lambda =\kappa $.

\begin{figure}[htbp]
\begin{center}
 \includegraphics[width=.45\linewidth,origin=tl]{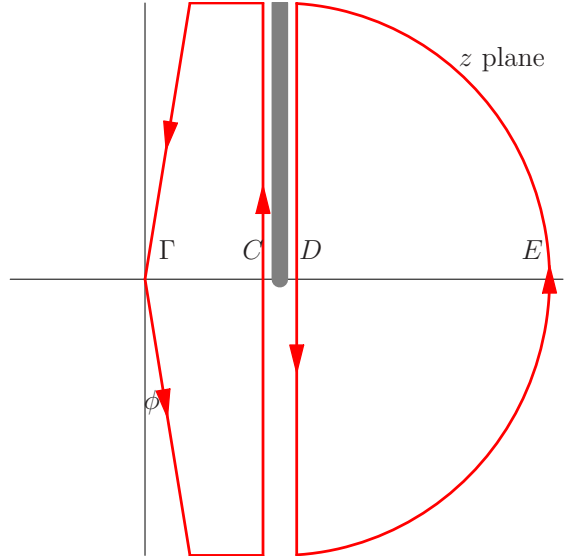}
\end{center}
\caption{Two contours avoiding the branch-cut.
} \label{fig:contour2}
\end{figure}

The presence of the branch-cut suggests
that one needs to include another mode
$\tilde p_0 = -\kappa \ln (z/(\kappa a) -1 )$,
which appears as a new real pole in Eq.~(\ref{eq:hm})
in the Feynman propagator (\ref{propagator}).
Suppose one consider two contour integrals,
I and II.
I consists of 4 components
in Fig.~\ref{fig:contour2},
$\Gamma$ along the imaginary axis
at $z=0$, $C$ along the branch-cut
at  $z=\kappa a -\epsilon + iy $,
and the rest at $z=\pm i \infty$ between $z=0$ and
$z=\kappa a $.
II consists of 2 components,
$D$ along the branch-cut
at  $z=\kappa a +\epsilon + iy $
and $E$ along the large half circle.

Contour integration I is defined as
\begin{eqnarray}
{\rm I}&=&\oint d z \,
    e^{-\sigma z + \frac{\alpha p_0(a;z)}{\kappa } }\,  p_0(a;z)
    \frac{d}{dz} \ln  f_l(z) \,,
\end{eqnarray}
where $ p_0 (a; z)
= -\kappa \ln \left(1 - {z}/{\kappa a } \right) $
and $\sigma$ is introduced to regularize the integral.
This integration is written as
\begin{equation}
\label{eq:C-relation}
{\rm I}_\Gamma + {\rm I}_{C}  =
\sum_{z_n < \kappa a}  p_0 (a; z_n)
       e^{-\sigma z_n+ {\alpha  p_0 (a;z_n)}/{\kappa } }\,,
\end{equation}
where $\rm{I}_\Gamma$ and  $\rm{I}_{C}$  denote the  integrations
along  segments $\Gamma$ and $C$, respectively.
Due to the regularization,
the integration along $z=\pm i \infty$ vanishes.

Contour integration $\rm{II}$ is defined as
\begin{eqnarray}
\label{eq:D-relation}
{\rm II}&=&\oint d z \,
    e^{-\sigma z + \frac{\alpha \tilde p_0(a;z)}{\kappa } }\,  \tilde p_0(a;z)
    \frac{d}{dz} \ln  f_l(z) \,,
\end{eqnarray}
where $ \tilde p_0 (a; z)
= -\kappa \ln \left( {z}/{\kappa a } -1 \right) $.
This integration gives the relation
\begin{equation}
{\rm II}  ={\rm II}_{D}  =
\sum_{z_n > \kappa a}  \tilde p_0 (a; z_n)
       e^{-\sigma z_n+ {\alpha  \tilde p_0 (a;z_n)}/{\kappa } }
\end{equation}
because $\rm{II}_E$ vanishes.

On the other hand, $\rm{I}_{C}$ and $\rm{II}_{D}$
are  written as
\begin{eqnarray}
{\rm I}_{C}  &=&-\kappa \int_{-\infty}^\infty \frac{dy}{2\pi}
e^{-\sigma (\kappa + i y)}
 \frac{ \ln \left (- \frac{iy}{\kappa a} \right) }
{ \left (- \frac{iy}{\kappa a} \right)^\alpha}
 \frac{d}{dz} \ln  f_l(z)\Big|_{z= \kappa+iy }\,, \\
{ \rm II}_{D}  &=&+ \kappa \int_{-\infty}^\infty \frac{dy}{2\pi}
e^{-\sigma (\kappa + i y)}
 \frac{ \ln \left ( \frac{iy}{\kappa a} \right) }
{ \left ( \frac{iy}{\kappa a} \right)^\alpha}
 \frac{d}{dz} \ln  f_l(z)\Big|_{z= \kappa+iy }\,.
\end{eqnarray}
One may have the following relation between the integrations:
\begin{eqnarray}
 {\rm II}_{D}  &=& -(-1)^\alpha \rm{I}_{C} + B \,,\\
 B&=&  \mp \kappa \pi (i) ^{\alpha +1} \int_{-\infty}^\infty \frac{dy}{2\pi}
e^{-\sigma (\kappa + i y)}
 \frac{ 1 }
{ \left ( \frac{y}{\kappa a} \right)^\alpha}
 \frac{d}{dz} \ln  f_l(z)\Big|_{z= \kappa+iy }\,,
\end{eqnarray}
where $B$ appears due to the branch-cut and
its sign $\mp$ depends on the branch-cut position,
which may lie either on the upper half plane or on the lower half plane.

When $\alpha=0$,
the branch-cut contribution, $B$,
can be understood if
one considers an integration from a discrete mode $z_n$:
\begin{eqnarray}
J&=&
\int_{-\infty}^\infty dy \,\,
\frac{e^{-\sigma z}} {z -z_n} \Big|_{z=\kappa +i y }
= e^{-\sigma \kappa}  \int_{0}^\infty
dy  \Re
\frac{e^{-i\sigma  y }} {\kappa+iy -z_n }  \nn\\
&=&  2e^{-\sigma \kappa}  \int_{0}^\infty
dy \frac {(\kappa-z_n) \cos (\sigma y) - y \sin (\sigma y) }
{(\kappa -z_n)^2 +y^2}.
\end{eqnarray}
Since $\cos (\sigma y)$ and  $\sin (\sigma y)$ are oscillating functions, one can put $\sigma \to 0$ before
the integration.
In this case, one can see that $J$ is real and
is evaluated as  $\pi$, which is independent of $z_n$.
Thus, each mode's contribution is independent of $a$ and
goes away when the contribution of the radius $\eta R$
is subtracted. Thus, $B$ is imaginary, but does not contribute to the Casimir energy when $\alpha=0$.

When $\alpha=3$, one has an integration from the branch-cut contribution:
\begin{eqnarray}
K &=&
\int_{-\infty}^\infty dy \left(
\frac{\kappa a}{y} \right)^3
\frac{e^{-\sigma z }} {z -z_n} \Big|_{z=\kappa +i y }
\\
&=& i e^{-\sigma \kappa}  \int_{0}^\infty dy \left(
\frac{\kappa a}{y} \right)^3
\frac {(\kappa-z_n) \sin (\sigma y) + y \cos (\sigma y) }
{(\kappa -z_n)^2 +y^2}
\nn\\
&\to& i \left(
\frac {\kappa a} {\kappa -z_n} \right)^3
 \int_{0}^\infty  \frac{dy}{y^2}
\frac 1 {1 +y^2 }
\nn.
\end{eqnarray}
In this case, $\sigma $ is not effective 
in regulating the theory,
and $K$ not only diverges due to the
singularity at $y=0$ but also depends on each discrete mode $z_n$.
This non-vanishing branch-cut contribution
makes $B$ real.
However, $B$ has a  sign ambiguity, which is not physical
because the real world should not
depend on the branch-cut's position.
To make the branch-cut independent, one may average
the branch-cut's contribution to get rid of the
branch-cut's arbitrariness. As a result, $B$ vanishes.

Finally, we are left with the relations
\begin{eqnarray}
 {\rm II}_{D}^{reg}  &=& -(-1)^\alpha {\rm I}_{C}^{reg}
\end{eqnarray}
and
\begin{eqnarray}
{\rm I}_{\Gamma} =\sum_{z_n < \kappa a}  p_0 (a; z_n)
       e^{{\alpha  p_0 (a;z_n)}/{\kappa } }
+
\sum_{z_n > \kappa a}   (-1)^\alpha \tilde p_0(a; z_n)
    e^{{\alpha  \tilde p_0 (a;z_n)}/{\kappa } } \,.
\end{eqnarray}
These expressions hint  that one needs to identify
$\omega_{\bf p}^{(3)}$ as
\begin{equation}
\label{energyid}
 \omega_{\bf p}^{(3)}  =  (-1)^\alpha  p_0^{(3)} \,, \end{equation}
where $\alpha=0$ or $3$ and
\begin{eqnarray}
E_c^{(P)}
&=&  \frac {\hbar} 2 \Big(  \int_{|{\bf p}|< \kappa}
 \frac{d^3 \bf p}{(2\pi)^3} \,\omega^{(+)}_{\bf p} \,
e^{\alpha \omega^{(+)}_{\bf p} /{\kappa} }
 +  \int_{|{\bf p}|> \kappa}  \frac{d^3 \bf p}{(2\pi)^3}
  \,  \omega^{(3)}_{\bf p} \,
e^{-\alpha \omega^{(3)}_{\bf p}/{\kappa} }
   \Big) \,.
\end{eqnarray}
This definition gives $\kappa \to  - \kappa$
symmetry.
Noting that $\rm{I}_\Gamma$ is the same as the one
obtained from the anti-particle contribution
in the contour integrations $\Gamma_1$ and $\Gamma_2$
in Fig.~\ref{fig:contour},
we have particle and anti-particle symmetry
of the vacuum $E_c^{(P)} = E_c^{(A)}$.
This demonstrates that the high-momentum
(HM) mode  $\omega^{(3)}_{\bf p}$, which
exists only when the momentum is greater than $\kappa$,
will make  the particle 
and the anti-particle contribution to the Casimir energy
equal.
In other words, the vacuum respects the particle and
the anti-particle symmetry.

However, this is in a serious contradiction
with the result from the thermal response calculation.
Even though the blackbody radiation~\cite{kimrimyee}
and the vacuum in an acceleration frame~\cite{KRY-field}
distinguish the particle and the anti-particle
responses at the order of $O(1/\kappa)$,
one needs to avoid the HM mode
because in the presence of
the high-momentum mode,
a particle with
a small energy may have a low
and a high momentum at the same time,
and HM can spoil the commutative result
because of the thermal density of the HM contribution 
which is the order of $O(\kappa^2 T^2)$~\cite{kimrimyee}
when $\kappa \to \infty$
($O(\kappa^2 T^2)$ if one uses the
relation in Eq.~(\ref{energyid})).
Thus, from the physics at the $\kappa \to \infty$ limit,
one naturally should avoid the HM mode.
This fact can also be seen in the dispersion relation
for the high-momentum mode in Eq.~(\ref{eq:hm}), 
which
does not have the proper commutative limit.
In this asymmetric ordering case, therefore,
the vacuum breaks the particle and the anti-particle
symmetry.

Thus, if one imposes the particle and the anti-particle
symmetry on the vacuum,
then one needs to modify the theory.
Even though the KMST is unique, the KPS is not:
Depending on the ordering of the kernel
of the Fourier transformation, the
$\kappa$-Poincar\'{e} algebra is differently realized.
There have been attempts~\cite{kosinski,kimrimyee,FGN,LRZ,amelino}
to construct the star-product of field theories.
If the exponential kernel function
of the Fourier transformation
is ordered symmetrically,
then the Casimir invariant given in 
Eq.~(\ref{eq:casimir-asm-inv})
changes into
\begin{equation}
M_s^2(p)=\left(2 \kappa \sinh
\frac{p_0}{2 \kappa }\right)^2- {\bf p}^2 \,,
\end{equation}
thus, the dispersion
relation of the massless field changes into
\begin{equation}
\omega_{\bf p}=2 \kappa \ln \left(\frac{|\bf p|}{2\kappa}
    +\sqrt{1+\frac{{\bf p}^2}{4\kappa ^2}} \,\right)
    =-2 \kappa \ln  \left(-\frac{|\bf p|}{2\kappa}
    +\sqrt{1+\frac{{\bf p}^2}{4\kappa ^2}} \,\right)
\end{equation}
instead of the one given in Eq.~(\ref{masslessmode}).

This massless mode has $\kappa \to - \kappa$
invariance and the particle and antiparticle dispersion
relation is simply $p_0 = \pm \omega_{\bf p}$.
In this case, because there is no branch-cut ambiguity,
one can expect the particle and the antiparticle symmetry
of the vacuum and may construct a field theory
with $\kappa$-deformed Poincar\'e symmetry
on the symmetric vacuum.

On the other hand, the HM mode
is also known to appear  
in  particle and anti-particle  spectra
and to spoil the $\kappa \to \infty$ 
limit~\cite{kimrimyee}.
Thus, one has to restrict the on-shell spectra and
construct the field theory based on this
observation.  This restriction can be imposed as
$M_s^2(p) \ge 0 $
($M_s^2(p) =0 $ for the massless case and
$M_s^2(p) > 0 $ for the massive case),
which respects
the $\kappa$-deformed Poincar\'e symmetry.
The details of this investigation
will be provided in a separate paper.

In addition to the vacuum symmetry,
the ordering is well known to affect 
the Casimir energy from the studies of 
Refs.~21-23 
for the case of two infinite parallel plates.
The deformed effect is seen at the order $1	/\kappa^2$,
but the contributions are drastically different.
The symmetric ordering deformation
gives a more attractive effect~\cite{bowes,cougo} whereas
the asymmetric ordering deformation reduces the attraction and can result in a stable configuration at a certain range of $\kappa a$~\cite{nam}.
The convergence of the $1/\kappa$ expansion
is considered in Ref.~22 
where  the $1/\kappa$ expansion might
turn out to be an asymptotic series expansion rather than
a converging series expansion.
It is not clear yet how these 
results~\cite{bowes,cougo,nam}  will change if the KPS 
measure is incorporated.
The structure of the higher-order series expansion is
to be studied carefully in this spherical geometry also
and is beyond the scope of this paper.

We remark in passing that the angular momentum summations of $B_1(\nu, a)$ and $B_1(\nu, \eta R)$ are finite
and are $O(\kappa)$, as seen in Eq.~(\ref{Ec-max}),
even though we put the regularization $\sigma \to 0$
before the integration and summation.
In the commutative limit,
however, the terms $B_1(\nu, a)$ and $B_1(\nu, \eta R)$
become infinity
and cannot be evaluated without a proper regularization.

Finally, suppose one considers the early Universe
and takes the Casimir energy as
one of the main radiation sources to the Universe
after the inflationary regime
because the excitation modes decay away,
but the Casimir energy is just the vacuum energy
and  might survive during the inflation.
Then, at the final regime of density fluctuations,
the Casimir energy may leave some effect
on the global structure of our Universe.
Note that the Casimir energy of a sphere 
measures the finite-size-corrected energy  
with respect to the infinite-size vacuum energy 
and is given as $O(1/a)$.
Therefore, the energy density 
inside the sphere is proportional to $O(1/a^4)$.
In addition, one can confirm that
most of the finite-size Casimir energy comes
from the lower part of the $l$ modes,
about 90\% of the contribution comes from $l=0$ to $4$.
The $l=0$ mode is the angular-independent contribution,
and the $l=1$ mode can be removed
by the motion of observer.
Therefore, the $l=2$ mode would be the most relevant mode
in the cosmological sense,
and it remains to be seen if
its $\kappa$-deformed correction can be
detected at the large scale of the present Universe.

\begin{appendix}
\section{Evaluation of the Divergent part ${\cal E}_0$} \label{sec:appA}

The regularized angular momentum mode of the Casimir energy
is given in Eq.~(\ref{eq:regE}):
\begin{eqnarray}
E_l^{reg} (a) &=& -\frac{\kappa\nu}{\pi}
 \Re \int_0^\infty dy\, e^{- i \sigma ye^{-i \phi}}
i g(a, i y e^{-i\phi}) \frac{d}{dy} \log \lambda_\nu (ye^{-i\phi})
\nn\\
&=& -\frac{\kappa\nu}{\pi}
 \Re \int_0^\infty dy\, e^{- i \sigma \nu ye^{-i \phi}}
i g(a, i \nu y e^{-i\phi}) \frac{d}{dy} \log \lambda_\nu (\nu ye^{-i\phi})\,,
\end{eqnarray}
where we rescale $y$ as $\nu y$ so that we can use the
the explicit large-order behavior of the Bessel function~\cite{abramwitz}.
For large $\nu$, the large-order behavior of $\lambda_\nu(\nu y)$ is given as
\begin{eqnarray}
\log \lambda_{\nu}(\nu y)
\equiv\log \Big(2 \nu y \,I_\nu(\nu y) K_\nu(\nu y)\Big) &=&
    \sum_{n=0}^\infty\frac{q_n(y)}{\nu^{2n}},
\end{eqnarray}
where
\begin{eqnarray} \label{eq:q12}
q_0(y) &=& \frac12 \log \frac{y^2}{1+y^2}, \\
q_1(y) &=& \frac{y^2}{8(1+y^2)^2}\left(1-\frac{5}{1+y^2} \right), \nn \\
q_2(y) &=& \frac{y^2}{64(1+y^2)^3}\left( 13-\frac{271}{1+y^2}
    +\frac{791}{(1+y^2)^2}- \frac{565}{(1+y^2)^3} \right),  \nn
\end{eqnarray}
and $q_{n\geq 1}(y)$ is $O(y^{-2n})$ for large $y$.
The Casimir energy is rewritten in terms of the large-order behavior as
\begin{eqnarray}
E_c(a)
&=& \sum_{n=0}^\infty \Big( {\cal E}_n (\sigma, a)
- {\cal E}_n (\sigma, \eta R) \Big) \,,\\
{\cal E}_n (\sigma, r)
&=&  - \sum_l  \frac{\kappa}{\pi\nu^{2n-1}}
\Re\int_0^\infty dy\,
i e^{- i \sigma \nu ye^{-i \phi}}\,
g(r, i \nu y e^{-i \phi})
\frac{d}{dy}  q_n(y e^{-i \phi})\,,
\nn
\end{eqnarray}
where the limits ${R \to \infty, \sigma \to 0, \phi \to 0}$
are to be taken at the end.

Let us consider ${\cal E}_{0}(\sigma, a)$ in detail.
${\cal E}_{0}(\sigma, a) $ is divergent
when $\sigma \to 0$ before summing over $l$.
Thus, one needs to evaluate this term with non-vanishing $\sigma$:
\begin{equation}
\label{eq:E0}
{\cal E}_{0}(\sigma, a) =
-\frac{\kappa}{\pi}\sum_{l=0}^\infty\Re \int_0^\infty dy ~y
    e^{-2i\phi}\nu  e^{-i \sigma \nu y e^{-i\phi}}
    i g(a, i  \nu y e^{-i\phi}) \frac{1}{(1+ y^2 e^{-2i\phi})(y^2 e^{-2i\phi})} \,.
\end{equation}
Formally, one can write
\begin{eqnarray}
{\cal E}_0 (\sigma, a)
&=& - \frac{\kappa}{\pi}
 \Re \,\, g(a, -\frac{\partial}{\partial \sigma} )\,
  \Big( -\frac{\partial}{\partial \sigma}\Big)\,
\int_0^\infty  {dy} e^{-i \phi}
 \sum_l e^{- i \sigma \nu ye^{-i \phi}}\,
 \frac{1}{(1+ y^2 e^{-2i\phi})(y^2 e^{-2i\phi})} \nn\\
&=& \frac{\kappa}{2\pi}
 \Re  g(a, -\frac{\partial}{\partial \sigma} )
  \Big( -\frac{\partial}{\partial \sigma}\Big)\,
\int_0^\infty  {dy}{ e^{-i \phi}}
  \frac {i}{ \sin (\frac{\sigma ye^{-i \phi}}2 )}\,
  \frac{1}{(1+ y^2 e^{-2i\phi})(y^2 e^{-2i\phi})}\,.
\end{eqnarray}
The integral can be done using a change of variable
$y \rightarrow y e^{i\phi} $ because 
the angular integral vanishes at $\infty$.
Then, the line integral is finite 
and becomes pure imaginary,
and the real part vanishes.
(One can be convinced 
that  the integration near $y =0$ is finite
from Eq.~(\ref{eq:E0}) directly.)
This allows one to ignore ${\cal E}_{0}(\sigma, a)$ and
${\cal E}_{0}(\sigma, \eta R)$ completely.

\section{  ${\cal E}_1 (a)$ and $ {\cal E}_2 (a)$ when $\alpha=0$ } \label{sec:appB}

$ {\cal E}_n (a)$ in Eq.~(\ref{eq:Enr})   for $\alpha=0$  is given as
\begin{eqnarray}
{\cal E}_n (r) &=& \frac1r \sum_l  \frac{B_n (\nu, r) }{\nu^{2n-2}}\,,
\qquad B_n (\nu, r)  = \frac1{\pi }
\int_0^\infty dy\, \frac{
    {q}_n(y)}{1+\frac{\nu^2 y^2}{(\kappa r)^2}}\,.
\end{eqnarray}
In this appendix, we evaluate ${\cal E}_1 (r)$ and ${\cal E}_2 (r)$
in two different ways.  One is to sum over $l$ first and
to evaluate the
integration later.  The other way is to integrate first and
to sum later.
Both ways provide useful viewpoints.

Let us consider
\begin{eqnarray}
{\cal E}_1 (r) &=& \frac1r \sum_l  {B_1 (\nu, r) } \,,
\qquad B_1 (\nu, r)  = \frac1{\pi }
\int_0^\infty dy\, \frac{
    {q}_1(y)}{1+\frac{\nu^2 y^2}{(\kappa r)^2}}\,.
\end{eqnarray}
Using the summation result
$$
\sum_l \frac{1}{1+\frac{\nu^2 y^2}{(\kappa r)^2}}
=\frac{\pi\kappa r}{2 y}  {\tanh (\pi \kappa r/y)}\,,
$$
one has
\begin{equation}
{\cal E}_1 (r) =\frac {\kappa}2 \int_0^\infty
\frac{dy}y q_1(y)
\tanh \left( \frac {\pi \kappa r}y \right)\,.
\end{equation}
This integration is not convergent 
and is subtracted
by ${\cal E}_1 (\eta R)$:
\begin{eqnarray}
{\cal E}_1 (a) - {\cal E}_1 (\eta R)
&=&\frac {\kappa}2 \int_0^\infty
\frac{dy}y q_1(y)
\left\{ \tanh \left( \frac {\pi \kappa a}y \right) -1 \right\}
\nn\\
&=& \frac {\kappa}{16}
\Big( \frac1{\pi \kappa a} \Big)^2
\int_0^\infty d\xi
\frac{\xi }{1 + (\frac{\xi}{\pi \kappa a })^2 }
\left( 1 - \frac{5 (\xi/\kappa \pi a)^2 }{1 + (\frac{\xi}{\pi \kappa a}
)^2} \right) ( \tanh \xi -1 )
\nn\\
&=&-\frac{1}{\kappa a^2} \frac 1{384} \Big(
1 - \frac{21}{20} \left( \frac1{\kappa a} \right)^2 +
O \left( \frac1{\kappa a} \right)^4 \Big)
\,,
\end{eqnarray}
where   $R \to \infty$ is taken.
It is to be noted that  the limiting procedure
is taken for the case $\kappa>0$. If one considers
the case with $\kappa<0$, one has to use the absolute
value of $\kappa$.

Likewise, for ${\cal E}_2 (r)$, one has
\begin{eqnarray}
{\cal E}_2 (r) &=& \frac1r \sum_l  \frac{B_2 (\nu, r) }{\nu^2 } \,,
\qquad B_2 (\nu, r)  = \frac1{\pi }
    \int_0^\infty dy\, \frac{
    {q}_2(y)}{1+\frac{\nu^2 y^2}{(\kappa r)^2}}\,.
\end{eqnarray}
The summation over $l$ gives
$$
\sum_l \frac{1}{\nu^2 \left(1+\frac{\nu^2 y^2}{(\kappa r)^2} \right)}
=\frac{\pi^2 }2  \Big( 1- \frac{y}{\kappa \pi r }
{\tanh \left( \frac{\pi \kappa r}{y} \right)} \Big)\,,
$$
and one has
\begin{equation}
{\cal E}_2 (r) =\frac {\pi}{2r} \int_0^\infty
\frac{dy}y q_2(y)
 \Big( 1- \frac{y}{\kappa \pi r }
{\tanh \left( \frac{\pi \kappa r}{y} \right)} \Big)\,.
\end{equation}
This integration is convergent, and ${\cal E}_2 (\eta R)$ vanishes
as $R \to \infty$:
\begin{eqnarray}
{\cal E}_2 (a)
&=&\frac {\pi}{2 a}
\int_0^\infty \frac {d\xi}{\xi^2 }
q_2( 1/\xi) \Big( 1 - \frac{\tanh \left( {\pi \kappa a \xi} \right)}{\kappa \pi a \xi }
 \Big)
\nn\\
&=& \frac{\pi^2}{2a} \frac{35}{32768}
+\frac{1}{\kappa a^2} \Big( \frac 1{256}
- \frac{13}{3072} \left( \frac1{\kappa a} \right)^2 +
O \left( \frac1{\kappa a} \right)^4 \Big)
\,.
\label{eq:E2a}
\end{eqnarray}

Now one may, instead, use integration first and  get
\begin{eqnarray}
B_1(\nu, a)&=& -\frac{1+\frac{11 \nu}{\kappa a}}{128
    (1+\frac{\nu}{\kappa a})^3 }  \,, \\
B_2(\nu, a )&=&
 \frac{35 +\frac{210 \nu}{\kappa a}+ \frac{562 \nu^2}{
    (\kappa a)^2}+\frac{4250\nu^3}{(\kappa a)^3}- \frac{657 \nu^4}{
    (\kappa a)^4}}{32768 (1+\frac{\nu}{\kappa a})^6 }\,.
\nn
\end{eqnarray}
Suppose one expands  $B_1(\nu,a )$ and $B_2(\nu,a )$
in $1/(\kappa a)$:
\begin{eqnarray} \label{eq:series:app}
B_1(\nu) &=& -\frac{1}{128}-\frac1{16}\frac{\nu}{\kappa a}+
    \frac{27}{128}\frac{\nu^2}{(\kappa a)^2} +\cdots , \nn \\
B_2(\nu)&=& \frac{35}{32768}
    +\frac{37}{32768} \frac{\nu^2}{(\kappa a)^2}
    +\cdots . \nn
\end{eqnarray}
In this expansion, there is no $1/(\kappa a)$ term in $B_2(\nu)$. However, as can be seen above in Eq.~(\ref{eq:E2a}), the summation of $B_2(\nu)/\nu^2$ over $l$ contains a $1/(\kappa a)$ term.  
This implies that
the naive series expansion in   $1/(\kappa a)$ is not valid.
One finds that there are nontrivial contributions at $\nu \sim \kappa a$
and that the large-order form of $l\sim \kappa a$
contributes to the summation to result in  $O(1/(\kappa a))$:
\begin{eqnarray}
{\cal E}_1 (a)-{\cal E}_1 (\eta R) &=& \frac1a
\sum_{l=0}^\infty \Big(B_1(\nu,a ) - B_1(\nu,\eta R )\Big)
\nn\\
 &=& -\frac{\kappa^2 a}{128}\left[ 11
    \psi^{(1)}(\frac{1}{2}+\kappa a) + 5\kappa a
    \psi^{(2)}(\frac{1}2+\kappa a) - \frac{\eta R}{a}
    (a\rightarrow \eta R) \right] \nn \\
  &=& -\frac{1}{384}\frac{1}{\kappa a^2 }
  \left(1+ O\left( \frac1 {\kappa a} \right)^2\right)\,,
\\
{\cal E}_2 (a) &=&
\frac 1 a \sum_{l=0}^\infty \frac{B_2(\nu)}{\nu^2} \nn\\
&=& \frac{1}{32768 a}
    \left[ \frac{35\pi^2}2 -35\psi^{(1)}(\frac{1}{2}+\kappa a)
    +35\kappa a\psi^{(2)}(\frac{1}{2}+\kappa a)
    \right.\nn\\
  &&\left.  -127(\kappa a)^2\psi^{(3)}(\frac{1}{2}+\kappa a)-226(\kappa
    a)^3\psi^{(4)}(\frac{1}{2}+\kappa a)-\frac{113}{3}(\kappa
    a)^4\psi^{(5)}(\frac{1}{2}+\kappa a) \right]\nn \\
  &=& + \frac1a\left( \frac{\pi^2}{2} \frac{35}{32768  } +
    \frac1{256} \frac1{\kappa a}
   + O\left(\frac1{\kappa a} \right)^3 \right)\,,
\end{eqnarray}
where $\psi^{(n)}(z)$ is the poly-Gamma function.

\section{$1/\kappa^2$ correction } \label{sec:appC}

In this appendix, we evaluate the dominant contribution
of $ \Delta {\cal E}_{n\ge 3} (a)$ to the Casimir energy:
\begin{eqnarray}
\label{app:Int:Ea}
\sum_{n \ge 3} \Delta  {\cal E}_n (a)
&=& \frac 1a \sum_{n \ge 3,\,l} \frac {\Delta  B_n (\nu, a) } {\nu^{2n-2}}
= \frac 1{\pi a} \sum_{n \ge 3,\,l} \frac {1 } {\nu^{2n-2}}
\int_0^\infty dy \, {q_n(y) } \Big(
{G\left(\frac{\nu y}{\kappa a}\right) - G(0)} \Big)\,,
\end{eqnarray}
where $G(0) =1$.
Noting that $q_n (y)= O (y^{-2n})$, one may divide the
integral as
\begin{eqnarray}
\label{integral-division}
\int_0^\infty dy \, {q_n(y) } \Big(
{G\left(\frac{\nu y}{\kappa a}\right) - G(0)} \Big)
&=& \int_0^\infty dy \, {q_n(y) }
\Big( G\left(\frac{\nu y}{\kappa a}\right)  -1
-  \left(\frac{\nu y}{\kappa a}\right)^2 \,  G_1\Big)
\nn\\
&&\qquad \qquad
+ \int_0^\infty dy \, {q_n(y) }
\Big(   \left(\frac{\nu y}{\kappa a}\right)^2
\,G_1 \Big)\,,
\end{eqnarray}
where  $G_1=\left.\frac12 \frac{d^2}{d y^2}G(y)\right|_{y=0}$.
We use the fact that 
the odd derivative of $G(y)$ at $y=0$ vanishes.
From this decomposition, one may put the summation as
\begin{eqnarray}
&&\sum_{n \ge 3} \Delta  {\cal E}_n (a)
= E_c ^{(2)} (a) + E_c ^{(3)} (a)\,,
\\
&&E_c ^{(2)} (a)
= \frac {G_1}{\pi a} \sum_{n \ge 3}\sum_{l} \frac {1 } {\nu^{2n-2}}
\int_0^\infty dy \, {q_n(y) }
 \left(\frac{\nu y}{\kappa a}\right)^2 \,,
\nn\\
&&E_c ^{(3)} (a)
= \frac 1{\pi a} \sum_{n \ge 3}\sum_l \frac {1 } {\nu^{2n-2}}
\int_0^\infty dy \, {q_n(y) }
\Big( G\left(\frac{\nu y}{\kappa a}\right)  -1
-  \left(\frac{\nu y}{\kappa a}\right)^2 \, G_1 \Big)\,,
\nn
\end{eqnarray}
whose contribution turns out to be convergent and is
order of $O(1/\kappa^2)$ and $O(1/\kappa^3)$, respectively.

\begin{table}[htbp]
\begin{center}
\begin{tabular}{|c|c|c|}
 \hline
  $l$ & $J(l)$ & $J_{\rm asymp}(l)$ \\
 \hline $0$& 0.00102501 & 0.00344 \\
 \hline $1$& $0.000287343$ & 0.000382667 \\
 \hline $2$& $0.000122372$ & 0.00013776 \\
 \hline $3$& $0.0000661443$ & 0.0000702857 \\
 \hline $4$& $0.0000410683$ & 0.0000425185 \\
 \hline $5$& $0.0000278738$ & 0.0000284628 \\
 \hline $6$& $0.000020120$ & 0.0000203787 \\
 \hline $7$& $0.0000151907$ & 0.0000153067 \\
 \hline $8$& $0.0000118677$ & 0.000011917 \\
 \hline $9$& $9.52338\times 10^{-6}$ & $9.54017\times 10^{-6}$ \\
 \hline $10$& $7.80878 \times 10^{-6}$ & $7.80952\times 10^{-6}$ \\
 \hline
\end{tabular}
\end{center}
\caption{Comparison of the values $J(l)$ 
with the corresponding values of the asymptotic  $J_{\rm asymp}(l)$.
 } \label{table1}
\end{table}

To evaluate $E_c ^{(2)} (a)$, one notes that
\begin{equation}
E_c ^{(2)} (a)
= \frac {G_1}{\pi a}\left(\frac{1}{\kappa a}\right)^2
\sum_{l \ge 0}  J(l)\,,
\end{equation}
where
\begin{eqnarray}
\label{app:J}
J(l) &\equiv& \nu^4
\int_0^\infty dy \,  y^2  \sum_{n \ge 3} \frac {q_n(y)  } {\nu^{2n}}
\\
&=& \nu^4 \int_0^\infty dy \, y^2
    \left[\log \lambda_\nu(\nu y)-q_0(y)-\frac{q_1(y)}{\nu^2}-
    \frac{q_2(y)}{\nu^4}\right] \nn.
\end{eqnarray}
The asymptotic form for large $l$, $l\gg 1$, is
proportional to $1/\nu^2\,$:
\begin{equation} \label{app:Jl}
J_{\rm asymp}(l)\simeq 0.000861 /(l+1/2)^2 \,.
\end{equation}
The numerical values of $J(l)$ are presented in Table~\ref{table1}.
Comparing this with  the value of $J_{\rm asymp}(l)$
for given $l$,
one notices that
$J(l)$ converges very fast to $J_{\rm asymp}(l)$.
Thus, we find that the summed value,
with the help of  $\sum_{l=0}^\infty (l+1/2)^{-2} =\pi^2/2$
for the large $l$ contribution as
\begin{eqnarray} \label{sum:Jl}
J_1=\sum_{l=0}^\infty J(l) \simeq 0.001713 ,
\end{eqnarray}
and we have $E_c^{(2)} $
\begin{equation}
\label{Ec-k2}
E_c^{(2)} = \frac{0.001713}{\pi a} \,\frac{G_1} {(\kappa a)^{2}}\,.
\end{equation}
Note that ${G_1}= -1$ when $\alpha=0$,
and  ${G_1}= -47/2 $ when $\alpha=3$.
\end{appendix}

\begin{acknowledgments}
This work was supported in part by a Korea Science and
Engineering Foundation grant (R01-2004-000-10526-0;
R\&Y), through the Center for Quantum Spacetime(CQUeST)
of Sogang University with grant (R11-2005-021; R),
and in part by a Korea Research Foundation grant funded by the
Korea Government (MOEHRD, Basic Research Promotion Fund
KRF-2005-075-C00009; K). R also thankfully acknowledges
the revision of this article done
during his visit to Korea Institute for Advanced Study.
\end{acknowledgments} \vspace{1cm}


\begin{thebibliography}{10}

\bibitem{casimir}
H. B. G. Casimir, Proc. K. Ned. Akad. Wet. {\bf 51}, 793 (1948).

\bibitem{boyer2}
T. H. Boyer, Phys. Rev. {\bf 174}, 1764 (1968); B. Davis, J. Math. Phys. {\bf 13}, 1324 (1972); R. Balian and B. Duplantier, Ann. Phys. (N.Y.) {\bf 112}, 165 (1978); K. A. Milton, L. L. DeRaad, Jr., and J. Schwinger, Ann. Phys. (N.Y.) {\bf 115}, 388 (1978).

\bibitem{milton}
M. Bordag, U. Mohideen, and V. M. Mostepanenko, Phys. Rep. {\bf 353}, 1 (2001);
K. A. Milton, J. Phys. A: Math. Gen. {\bf 37}, R209 (2004)
and references therein.

\bibitem{chen}
F. Chen, G. L. Klimchitskaya, U. Mohideen, and V. M. Mostepanenko, Phys. Rev. {\bf A69}, 022117 (2004).

\bibitem{autumn}
K.\ Autumn, M.\ Sitti, Y.\ A.\ Liang,
A.\ Peattie, W.\ W.\ Hansen, S.\ Sponberg,  T.\ W.\ Kenny, 
R.\ Fearing, J.\ N.\ Israelachvili, and R. J. Full, Proc. Natl. Acad. Sci. USA {\bf 99} 12252 (2002); K. Autumn, Am. Sci. {\bf 94}, 124 (2006).

\bibitem{boyers}
T. H. Boyer, Phys. Rev. {\bf D21}, 2137 (1980); D. W. Sciama, P. Candelas, and D. Deutsch, Adv. Phys. {\bf 30}, 327 (1981); S. Hacyan, A. Sarmiento, G. Cocho, and F. Soto, Phys. Rev. {\bf D32}, 914 (1985).

\bibitem{garriga}
J. Garriga, O. Pujolas, and T. Tanaka, Nucl. Phys. {\bf B605}, 4922 (2001): J. Garriga and A. Pomarol, 
Phys. Lett. {\bf B560}, 91  (2003).

\bibitem{kappaP}
J. Lukierski, A. Nowicki, H. Ruegg, and V. N. Tolstoy,
Phys. Lett. {\bf B264}, 331 (1991);
S. Majid and H. Ruegg, Phys. Lett. {\bf B329}, 189 (1994);

\bibitem{doubly}
G. Amelino-Camelia, 
Phys. Lett. {\bf B510}, 255 (2001);
Int. J. Mod. Phys. {\bf D11}, 35  (2002).

\bibitem{doubly-kappa}
N. R. Bruno, G. Amelino-Camelia, and J. Kowalski-Glikman,
Phys. Lett. {\bf B522}, 133 (2001);
J. Kowalski-Glikman and S. Nowak, Phys. Lett. {\bf B539}. 126 (2002)

\bibitem{majid}
S. Majid and H. Ruegg,
Phys. Lett. {\bf B334}, 348 (1994).


\bibitem{sitarz}
A. Sitarz, Phys. Lett. {\bf B349}, 42 (1995); C. Gomera, P. Kosi\'{n}ski, and P. Ma\'{s}lanka, J. Math. Phys. {\bf 37}, 5820 (1996).

\bibitem{gonera}
C. Gonera, P. Kosi\'{n}ski, and P. Ma\'{s}lanka, J. Math. Phys. {\bf
37}, 5820 (1996).


\bibitem{kosinski}
P. Kosi\'{n}ski, J. Lukierski, and P. Ma\'{s}lanka, Phys. Rev. {\bf D62}, 025004 (2000).

\bibitem{KRY-field}
 H.-C. Kim. J. H. Yee, and C. Rim, Phys. Rev. {\bf D75}, 045017 (2007).

\bibitem{kim}
H.-C. Kim, J. H. Yee, and C. Rim, Phys. Rev. {\bf D72}, 103523 (2005).

\bibitem{kimrimyee}
H.-C. Kim, C. Rim, and J. H. Yee,
Phys. Rev. {\bf D76}, 105012 (2007).

\bibitem{DLW}
M. Daszkiewicz, J. Lukierski and M. Woronowicz,
arXiv:0708.1561[hep-th].

\bibitem{gauge}
P. Aschieri, C. Blohmann, M. Dimitrijevic, F. Meyer, P. Schupp, and J. Wess,
Class. Quant. Grav. {\bf 22}, 3511 (2005).

\bibitem{FGN}
L. Freidel, J. Kowalski-Glikman, and S. Nowak,
Phys. Lett. {\bf B648}, 70 (2007).

\bibitem{nam}
S. Nam, H. Park, and Y. Seo, J. Korean. Phys. Soc. {\bf 42}, 467 (2003).

\bibitem{bowes}
J. P. Bowes and P. D. Jarvis, arXiv:gr-qc/9602016.

\bibitem{cougo}
M. V. Cougo-Pinto, C. Farina, and J. F. M. Mendes, Phys. Lett. {\bf B529}, 256 (2002).

\bibitem{nesterenko}
V. V. Nesterenko and I. G. Pirozhenko, Phys. Rev. {\bf D 57}, 1284 (1998).

\bibitem{hagen}
M. E. Bowers and C. R. Hagen, Phys. Rev. {\bf D 59}, 025007 (1998).

\bibitem{abramwitz}
M. Abramwitz and I. A. Stegun, {\it Handbook of Mathematical Functions} (National Bureau of Standards, Washington, D.C., 1964).

\bibitem{scalar-sphere}
C. M. Bender and K. A. Milton, Phys. Rev. {\bf D 50}, 6547 (1994);
A. Romeo, Phys. Rev. {\bf D 52}, 7308 (1995).

\bibitem{LRZ}
J.\ Lukierski, H.\ Ruegg, and W.\ Zakrzewski,
Ann.\ Phys.\ {\bf 243}, 90-116 (1995).

\bibitem{amelino}
A. Agostini, G Amelino-Camelia, and F. D'Andrea, Int. J. Mod. Phys. {\bf A 19}, 5187 (2004);
A. Agostini, F. Lizzi, and A. Zampini,
Mod. Phys. Lett. {\bf A 17}, 2105 (2002).

\end{thebibliography}
\end{document}